\documentclass[twocolumn,twoside,slac_two]{revtex4}
\usepackage{graphicx}
\usepackage{fancyhdr}
\begin{document}

\title{Hadron Spectroscopy -- Theory}

%

\author{E.S. Swanson}
\affiliation{Department of Physics and Astronomy, University of Pittsburgh, Pittsburgh PA 15260}

\begin{abstract}
A brief review of theoretical progress in hadron spectroscopy and nonperturbative QCD
is presented. Attention is focussed on recent lattice gauge theory, the Dyson-Schwinger formalism,
effective field theory, unquenching constituent models, and some beyond the Standard Model physics.
\end{abstract}

\maketitle

\thispagestyle{fancy}


\section{Introduction}

There are two, similarly prevalent, ideas about hadrons. The first maintains that they represent
`irreducible complexity' and that next to nothing can be learned by studying them. The second
holds that hadrons are simple and understood, and next to nothing can be learned by studying
them. Both can't be right and one might reasonably suspect that the true situation lies
somewhere in the middle. And this implies that profound things can indeed be learned about
the strongly interacting sector of the Standard Model by studying hadrons. In fact, many scientists are
motivated to study QCD to deepen 
our understanding of the emergent phenomena associated with nonperturbative field theory. These
phenomena include confinement, chiral symmetry breaking, topological excitations, deconfinement,
novel phases of matter, nonperturbative properties of glue, and many other topics.

The GlueX collaboration\cite{gluex} provides a good example of this ethos; the goal of this group
is to replicate the early successes in building quantum mechanics (via atomic spectroscopy) in
the gluonic sector of QCD. Specifically, if a gluonic spectroscopy can be discovered and decoded,
much can be learned about this intriguing, and largely unknown, sector of the Standard Model.

Unfortunately, the phenomenally successful methods for computing in field theory developed 
in the last century are generally useless when applied to low energy QCD. This has generated
something of a crisis in the field, and there is a popular notion that `nonperturbative' means
`noncomputable'. Fortunately, not everyone has given up. Lattice gauge theory has developed into
a powerful tool in the past twenty years. 
Other tools of varying pedigree are potential models, effective
field theory, potential models, the Schwinger-Dyson formalism, and a collection of ideas
centred on the operator product or multipole expansion.

These tools are being applied to a number of related issues. Among these are
vacuum structure (chiral symmetry breaking, confinement, instantons, vortices,
monopoles, chiral restoration); 
long range interactions (pomeron exchange, pion exchange, gluonic multipoles,
coupled channel effects, confinement, the emergence of nuclear physics); 
short range interactions (gluon exchange, pion exchange, instantons, coupled channels);
gluonics (hybrids, glueballs, strong decays); quark matter (phase transitions, quarkyonic
matter, small $x$ physics); and nucleon structure (generalised parton distributions,
transverse momentum distributions, single spin asymmetry, duality). A brief overview
of some of these topics follows.

\section{Nonperturbative Tools: Recent Results}

\subsection{Lattice}

A combination of Moore's Law and continually improving algorithm technique has pushed the
lattice gauge theory program into the truly useful regime. Lattices are now large, quark
masses are light, or nearly so, and virtual quarks are often incorporated into computations.
Furthermore, there has been substantial progress in constructing large bases of interpolating operators\cite{L-ops}, in smearing techniques\cite{L-smear}, in maintaining chiral symmetry\cite{L-chiral}, and in implementing all-to-all quark inversion algorithms\cite{L-many}. 

The combination of many of these advances now permits studies of excited states for the first time\cite{L-excited}; this is an important development that takes lattice gauge theory from the attractive, but nearly useless, enterprise of computing `gold-plated' quantities, to computing useful, but messy, observables\cite{L-abinitio}.

Unfortunately, a difficult problem lingers in this neighbourhood, namely how does one deal with
unstable states in a Euclidean regularisation of field theory? Luescher has suggested a method 
to extract scattering phase shifts\cite{L-luescher}, and this has been applied in preliminary fashion to the $\pi-\pi$ scattering length\cite{L-pipi}. Luescher's method applies to single-channel
elastic scattering, unfortunately, the crucial real-world generalisation to multi-channel inelastic scattering
remains to be made.

Deriving the main features of nuclear physics from QCD is a longstanding and challenging goal.
A few groups have recently begun examining the feasibility of obtaining nuclear interactions and
nucleon properties on the lattice and this effort now forms a substantial portion of the intellectual effort of the USQCD umbrella collaboration. An example nascent computation examines the interactions of five pions  in a 2.5 fm box\cite{L-nuclear}. The large nucleon masses and required high precision energies make this an {\it extremely} difficult problem. Reliable computations must await exoscale computers. An alternative approach is to use lattice techniques to solve a nucleon effective field theory\cite{L-Neff}; this can be useful, but of course only indirectly advances the goal of developing  nuclear physics from QCD.

Although somewhat removed from spectroscopy, lattice computations are also making important 
contributions to understanding RHIC experimental results and the structure of the quark and gluon matter at finite temperature and density.  A central result is the nuclear equation of state, which
is now being computed with (nearly) physical $u$, $d$, and $s$ quark masses\cite{L-eos}. These
results form a vital input to hydrodynamic models of RHIC scattering experiments. Unfortunately, they are less useful for supernova explosion models since these require an equation of state at low
temperature and moderate density, and lattice computations have a difficult time obtaining 
reliable results at finite density due to the fermion sign problem. It is also possible to 
obtain bulk transport properties of quark matter from the stress tensor correlation function\cite{L-bulk}, which are important to understanding RHIC experiment.

Recent lattice computations are also making significant contributions to heavy meson decays. 
For example, $B$ and $D$ decay form factors, decay constants,  and $B$ mixing parameters are now computed in unquenched LGT with light quarks. The results are crucial to the accurate extraction of CKM matrix elements\cite{L-ckm}
and, hence, of examining physics beyond the Standard Model (BSM). An interesting aside in this area
concerns the $D_s$ decay constant, which appears to be 4$\sigma$ larger than LGT predicts, and
raises the possibility that BSM (a $\tilde d$ leptoquark or a second Higgs doublet) contributes\cite{L-fds}.

Lattice BSM applications are not restricted to the $D_s$ decay constant. For example, an important
possibility for completing the Higgs sector of the Standard Model invokes dynamical electroweak
symmetry breaking (DEWSB). In the old days, this was postulated to occur via a TeV scale copy of QCD, dubbed technicolor (TC). TC rather neatly gives mass to the electroweak bosons, but has nothing
to say about fermion masses (or how non-eaten TC Goldstone bosons become massive). This problem can 
be overcome by considering additional gauge interactions (extended technicolor, ETC), with a symmetry that is broken at a higher scale. Integrating out the higher scale physics gives rise to four-quark operators that solve the aforementioned problems but introduce another one: large flavour changing neutral interactions\cite{BSM-etc}. This problem, in turn, can be possibly resolved if the ETC coupling runs sufficiently slowly. This scenario goes by the name `walking technicolor'\cite{BSM-wtc}.
It is thus important to determine parameters of QCD-like theories that permit the walking scenario\cite{BSM-luty}, specifically determining the number of flavours and colours that place the theory
near a nontrivial infrared fixed point. This task is ideally suited to lattice techniques\cite{BSM-fleming}.

\subsection{Schwinger-Dyson}

The Schwinger-Dyson equations form an alternative representation of a quantum field theory.
Truncating these equations then provides a method to obtain nonperturbative information about
the theory. Steady progress has been made in the past decade, starting with quite accurate 
computations of light meson properties using a simple model of the quark-gluon 
vertex\cite{SD-old}. Another important thread concerned the establishment of a viable 
mechanism for confinement. An early demonstration was made in Landau gauge, where it was
shown that infrared enhancement of the ghost propagator fulfilled the Kugo-Ojima confinement
criterion\cite{SD-conf}.  An accompanying advance involves the analysis of Schwinger-Dyson
equations in the infrared limit\cite{SD-review}; and this method can actually be applied
to the infinite tower of Schwinger-Dyson equations\cite{SD-all}.  

The connection to quark confinement can also be established in this approach\cite{SD-quark},
although more detailed studies are required before the community will comfortably declare
victory over the confinement problem.

The dominant issue in the Schwinger-Dyson formalism is how to reliably truncate the equations
while maintaining theoretically desirable features (such as the Ward-Takahashi identities).
It has been traditional to employ the rainbow-ladder truncation for, e.g., dynamical chiral
symmetry breaking, pseudoscalar mesons, and vector mesons. However this truncation is known
to be inadequate for axial-vector and scalar mesons\cite{SD-krass}. The former has been recently addressed 
by finding a Ward-Takahashi identity for the axial-vector Bethe-Salpeter equation with a fully
dressed quark-gluon vertex. Solving the resulting equation shows that chiral symmetry breaking
enhances spin-orbit splitting in the meson spectrum\cite{SD-axial}.

The method has also been extended to finite temperature and density, where it is especially useful
 for studying the chiral phase transition and properties of light mesons in-medium\cite{SD-finiteT}.
Of course it is difficult to study the interplay of chiral restoration and deconfinement in this
approach when quark-gluon vertex models are employed. 

A major missing element in the approach is reliable computation or modelling of the quark
interaction. Of course this should be confining, but it would be preferable if it contained
other well-known phenomenology of confinement, such as N-ality, the flux-tube limit, and Casimir
scaling\cite{SD-prop}.

\subsection{Modelling}

Nonrelativistic constituent models of low energy QCD predate QCD and are likely to be useful for
many years to come. Indeed, the constituent quark model continues to serve as a template for the 
discussion
of the properties of new states. Recent effort in this area has focussed on understanding the
new charmonium being discovered at the B factories, CLEO, FNAL, and BES (along with 
open charm $D_s$ mesons, heavy baryons, and a variety of light mesons). Because many of these
states lie high in their respective spectra, it is expected that coupling to nearby continuum
channels can affect their properties. In fact, it is speculated that some states are entirely
generated by such coupling\cite{M-new,M-rev}. Thus there is strong incentive to reliably model
the coupling of constituent $q\bar q$ states with the meson-meson continuum. This is not an
easy task, first one must understand $q\bar q$ creation, which is certainly a
nonperturbative gluonic process. Then one must find a way to compute the effects of mixing
with an infinite set of possible real and virtual decay channels. The result will heavily
shift the predictions of the `quenched quark model', and hence the model must be renormalised.
Indeed, one expects that virtual loop effects should disrupt naive valence expectations.
For example, the near degeneracy of the $\omega$ and $\rho$ mesons is hard to understand
in light of the different continuua that couple to these particles. The earliest observation
of this problem that I am aware of is called the `Oakes-Yang problem', after Oakes and 
Yang\cite{M-oy}
who noted that the Gell-Mann--Okubo mass formula should be ruined by virtual threshold effects.

There has been much work on this issue over the years. Strong decay models have been heavily 
investigated for several decades\cite{M-decays}, and are quite successful phenomenologically.
Studies of unquenching the quark model have also been performed in a variety of frameworks,
typically by integrating out the bound state or continuum channels in the coupled channel
Bethe-Heitler equations\cite{M-coupled}. The issue of renormalising the quark model was
raised and studied with a simple model in Ref. \cite{M-renorm}.

Recently a set of theorems concerning perturbative mass shifts and mixing
amplitudes of mesons due to meson loops were derived\cite{M-thm}. These theorems hold assuming that the quark pair
creation operator factorises into spin and spatial components and that the virtual mesons
fall into degenerate orbital multiplets. Under these conditions

(i) The loop mass shifts are identical for all states within a given N,L multiplet.

(ii) These states have the same total open-flavor decay widths.

(iii) Loop-induced valence configuration
mixing vanishes provided that $L_i \neq L_f$ or $S_i \neq S_f$.

These results have important implications: if the conditions are approximately true it implies
that meson mass shifts can be largely absorbed into a constant term in the quark potential.
Furthermore, if the virtual channels have masses split by subleading operators, then the
mesons that couple to them will follow the same patterns of mass shifts. This is a crucial
observation because it implies
that the quark model is robust under coupling to virtual meson channels.

I also note that mixing with the continuum can induce an effective short range interaction.
This interaction is in general spin-dependent and can have a substantial effect
on phenomenology. Sorting out spin-dependence due to gluon exchange, instantons,
relativistic dynamics, and continuum mixing remains a serious challenge for
model builders.

Finally, there is a lore that passing the lowest
continuum threshold in a given channel is associated with a deterioration of the
quality of potential models of hadrons. But hadron masses are
shifted throughout the spectrum (including below threshold) due to a given channel.
Thus, it is
the proximity of a continuum channel which can cause local distortions of the
spectrum. Of course, the problem in hadronic physics is that continuum
channels tend to get dense above threshold.

A number of issues remain open:
(a) One expects that when the continuum virtuality is much
greater than $\Lambda_{QCD}$ quark-hadron duality will be applicable and the sum
over hadronic channels should evolve into perturbative quark loop corrections to
the quark model potential. Correctly incorporating this into constituent
quark models requires marrying QCD renormalisation with effective models
and is not a simple task.
(b) Pion and multipion loops can be expected to dominate the virtual continuum
component of hadronic states (where allowed) due to the light pion mass. This raises
the issue of correctly incorporating chiral dynamics into unquenched quark models.
(c) Unquenching implies that simple constituent quark
models must become progressively less accurate high in the spectrum. However, this
effect is likely to be overwhelmed by more serious problems: the nonrelativistic
constituent quark
model must eventually fail as gluonic degrees of freedom are activated and because
chiral symmetry breaking (which is the pedestal upon which the constituent quark
model rests) becomes irrelevant for highly excited states\cite{M-restore}.
It is clear that an exploration of the excited hadron spectrum is required to
understand the interesting physics behind the unquenched quark model, gluonic
degrees of freedom, and chiral symmetry breaking.

The $X(3872)$ remains the poster child for coupled channel effect. In this regard it is interesting
to note that
there is increasing evidence that the $X$ must contain a substantial $c\bar c$ component. 
For example, the recently measured transition $X \to \gamma\psi'$ is comparable\cite{M-psi'} to the rate
for $X \to \gamma J/\psi$, which is difficult to accommodate in a molecular scenario. This is because the $\gamma J/\psi$ mode can come directly from the $\rho J/\psi$ or $\omega J/\psi$ component
of the $X$, while the $\gamma \psi'$ mode must come from the $D\bar D^*$ component with additional
excitation of the $c\bar c$ pair. Additional evidence for $c \bar c$ content comes from the $B$
decay branching fraction, which is typical for charmonium, and the ratio of $X$ production from
neutral and charged $B$ decay. It is therefore encouraging that direct computations show that the relative nearness of the $\chi'_{c1}$ can lead to a surprisingly large mixing between these states\cite{M-rev}.
Unfortunately, a rather severe problem hinders the implementation of this idea, namely it appears
impossible to develop a large hidden $c\bar c$ component of the $X$ without shifting its mass 
substantially lower thereby ruining its weak binding character. The only way the $X$ can remain
close to $D_0\bar D_0^*$ threshold and have a large $c\bar c$ component is if the mixing operator
is also of a weak binding nature. Why this should be true for an operator rooted in nonperturbative gluodynamics is difficult to see.

\subsection{Effective Field Theory}

Effective field theories are designed to simplify the computation of processes that are
dominated by several energy scales. They achieve this by implementing systematic power counting,
by summing logarithms of ratios of mass scales, and by taking advantage of QCD factorisation, amongst other things. The first effective field theory was chiral perturbation theory, and these days
the Standard Model is regarded as an effective field theory of a grander theory.

Many effective field theories exist depending on the dynamical system under consideration.
For example, 
soft collinear effective field theory (SCET) is applicable to highly energetic particles moving 
close to the light cone that interact with soft gluons\cite{EFT-scetOld}. Recent applications
of this theory to 
threshold enhancement in Drell-Yan processes, resummation of Higgs production at
hadronic colliders, and event shapes in $e^+e^-$ annihilation are reviewed in Ref. \cite{EFT-scet}.

Voloshin and Braaten pioneered the application of effective field theory to weakly bound hadrons
(with application to the $(3872)$\cite{EFT-X}. Here the idea is to  leverage the weak and nonrelativistic nature of the weakly bound states to extract universal features of such states. Further 
progress can be obtained by explicitly constructing the effective field theory in terms of 
$D$ and $D^*$ mesons\cite{EFT-Xfull}. For example, at leading order Fleming and Mehen obtain\cite{EFT-Xfull}

\begin{eqnarray}
&&\Gamma(X \to \chi_{c0}\pi^0) : \Gamma(X \to \chi_{c1}\pi^0) : \Gamma(X \to \chi_{c2}\pi^0)  = 
\nonumber \\
&& 4.76 : 1.57 : 1.0
\end{eqnarray}

Braaten and students have computed many properties of the $X(3872)$ in this formalism. Here I 
draw attention to a recent analysis of the decay of the $X$ to the $J/\psi \pi\pi$ and $D_0 \bar D_0 \pi^0$ modes. Their method accounts for the universal line shape expected for weakly bound states
and finds a consistent weakly bound state in both channels\cite{EFT-braaten}. This is contradiction
to other analyses that force the $X$ to resonate {\it above} threshold, which have found their
way into the PDG. Braaten points out that this has led to some authors misinterpreting the $X$ as
a virtual state. This issue remains controversial, for a more complete discussion see Ref. \cite{EFT-yulia}.

The final EFT considered here is potential NRQCD (pNRQCD). This is the application of the ideas of nonrelativistic QCD (NRQCD)\cite{EFT-nrqcd} (where scales of order $m_q$ are integrated out) to the regime below $m_q v$. In this case the degrees of freedom are singlet and octet fields and potentials. A novel
feature of this approach is that the potentials can be regarded as $r$-dependent matching
coefficients that appear upon integrating out scales of order the bound state momentum 
$m_q v \sim 1/r$\cite{EFT-pnrqcd}.
 
Although the formalism has been applied to many observables, I will restrict attention to
a computation of bottomonium  hyperfine splitting. This is of interest because this splitting
can allow a competitive determination of the strong coupling at this scale, the bottom quark mass,
possible BSM effects, and it serves to test EFTs. Of course this all depends on recent 
tour-de-force measurements of the $\eta_b$ mass\cite{EFT-etab}; the average mass obtained from 
$\Upsilon(3S) \to \eta_b \gamma$ and $\Upsilon(2S) \to \eta_b\gamma$ is 9390.4 $\pm 3$ MeV yielding
an experimental hyperfine splitting of

\begin{equation}
M(\Upsilon) - M(\eta_b) \equiv M_{hf} = 69.9 \pm 3\ {\rm MeV}.
\end{equation}

The splitting has been computed in pNRQCD employing masses accurate to ${\cal O}(m_b \alpha_S^5)$
that is resummed using the nonrelativistic renormalisation group. The result is weakly scale-dependent and is approximately\cite{EFT-penin}

\begin{equation}
M_{hf} = 39 \pm 11 \pm 10\ {\rm MeV}
\end{equation}
where the last error is due to uncertainty in $\alpha_S(M_Z)$. 
This prediction can be compared to an old quark model
prediction of $M_{hf} \approx 9460 - 9400 = 60$ MeV\cite{EFT-gi}, a simple quark model
prediction of $M_{hf} = 77$ MeV, and an unquenched lattice result of $M_{hf} = 61 \pm 14$ MeV\cite{EFT-lattice}.  The latter is of interest because the naive (and rough) agreement with experiment 
may be misleading. As Penin points out,  the lattice computation is made with a lattice spacing
$a \sim (\alpha_S m_b)^{-1}$ and thus misses hard contributions to the splitting. This contribution
can be estimated to be

\begin{equation}
\delta M_{hf} = -\frac{\alpha_S}{\pi}\frac{7 C_A}{4} \log(a m_b) M_{hf} \approx -20\ {\rm MeV}.
\end{equation}
Oddly, this brings the lattice result into agreement with the pNRQCD result, and both into
disagreement with experiment. The curiousness of the situation is only heightened by the good
agreement between experiment and pNRQCD that is found for charmonium.

\subsection{BSM Again}

It is worth pointing out that hadrons do not merely act as a nuisance in searches for BSM physics.
As we have already seen, QCD serves as an experimentally accessible paradigm for ETC models of
DEWSB. Thus theoretical techniques invented for QCD can possibly prove useful to future 
computations of, and in, grander theories. 

On the experimental side, hadrons can act as a source for BSM observables. Examples include
the measurement of CP and lepton flavour violation in $J/\psi$ decays, the effects of a
light pseudoscalar Higgs in $\Upsilon$ decays and in the hyperfine splitting, the effect
of leptoquarks on the $D_s$ decay constant (discussed above), the effect of nonstandard
Higgs-mediated leptonic decays of the $\Upsilon$, and the measurement of parity violating
electron scattering off the proton at low $Q^2$. The latter is being studied by the Qweak experiment at JLab,
which aims to the measure the proton's weak charge at approximately $Q^2 = 0.03$ GeV$^2$.

\section{Conclusions}

The B factories, BES, CLEO, and FNAL continue to discover new hadronic states and measure new
properties of these states. One fully expects this to continue with PANDA at FAIR, LHCb, the hadronic programs at ATLAS and CMS, new experiments to begin at JLab at 12 GeV, BESIII, and possibly SuperB at KEK. 

Interpreting these experimental results will rely on developing a robust understanding of the
properties of QCD in the strong regime. The substantial progress made in a variety of approaches to this 
problem has been outlined here. It is difficult not to be heartened by this progress. Furthermore, one 
fully expects it to continue into the foreseeable future.

\begin{acknowledgments}
I am grateful to the Department of Energy for support provided under contract
DE-FG02-00ER41135 and to E. Braaten, J. Dudek, C. Morningstar, C. Roberts, and  J. Terning
for illuminating discussions.
\end{acknowledgments}

\bigskip 


\begin{thebibliography}{99}
\bibitem{BSM-luty}
``The usual approaches are perturbative, so we know they are wrong. At least with technicolor, we
do not know it is wrong from the start". Markus Luty, private communication.
\end{thebibliography}
\end{document}